# Two strong nonlinearity regimes in cold molecule formation

A.M. Ishkhanyan[1], B. Joulakian[2] and K.-A. Suominen[3]

[1]*Institute for Physical Research NAS of Armenia, 0203 Ashtarak-2, Armenia*
[2]*Laboratoire de Physique Moléculaire et des Collisions, Université Paul Verlaine-Metz, 1 Bld Arago, 57078 Metz Cedex 3, France*
[3] *Department of Physics, University of Turku, 20014 Turun yliopisto, Finland*

*Two distinct strongly non-linear scenarios of molecule formation in an atomic Bose-Einstein condensate (either by photoassociation or Feshbach resonance) corresponding to large and small field detuning are revealed. By examining arbitrary external field configurations, we show that the association process in the first case is almost non-oscillatory in time while in the second case the evolution of the system displays strongly pronounced Rabi-type oscillations. We construct highly accurate approximate solutions for both limit cases. We show that at strong coupling limit the non-crossing models are able to provide conversion of no more than one third of the initial atomic population. Finally, we show that for constant-amplitude models involving a finite final detuning the strong interaction limit is not optimal for molecule formation.*

During the recent years, photoassociation [1], a process that occurs when two ultracold atoms collide in the presence of an optical field, and the Feshbach resonance [2], an analogous pairing process occurring in the presence of a magnetic field, became widely used tools for molecule formation in ultracold atomic ensembles. In particular, these techniques gained extensive attention as versatile routes for production of molecular Bose-Einstein condensates starting from atomic condensates [3] or ultra-cold degenerate Fermi gases [4].

The basic process for effective converting of atoms to molecules by these two techniques is to adjust a resonance, either photoassociative or Feshbach, using external time-dependent optical or magnetic fields. In this context, it is commonly believed that the strong field intensity regime is better for high conversion efficiency. This supposition emerged from the Landau-Zener analysis [5] widely used to model level crossing processes. However, the accumulated knowledge concerning the effective control of nonadiabatic quantum transitions by pulse shape/detuning effects in linear quantum systems (see, e.g., [6-11] and references therein) suggests that the Landau-Zener representations are substantially altered when more realistic models are discussed. Therefore, one would expect that in the nonlinear case of ultracold atomic photoassociation or Feshbach resonance the changes in the interaction picture caused by pulse shape or detuning modulations of the form other than the Landau-Zener model may be even more dramatic.

To address this question, we examine below the level crossing processes in general, i.e., assuming arbitrary pulse-shape and frequency-detuning time-configurations of the field.

We model the molecule formation process in a many-body quantum degenerate gas by a basic semiclassical nonlinear two-state mean-field model [12-13]. To reveal the most pronounced effects due to the nonlinearity, we focus on the case of strong interaction when the nonlinearity is a dominant feature. We arrive at several conclusions of general character.

First, we show that no non-crossing model is capable to provide conversion of more than ***one third*** of the atomic population. Thus, indeed, the tuning through the resonance is crucial for molecule production efficiency. Interestingly, we show that the molecular state probability is always very close to $1/6$ at the very point of the resonance crossing.

Furthermore, discussing the possible general limit cases of molecule formation under the strong nonlinearity conditions, we show that two qualitatively distinct regimes are possible. In the first case, corresponding to the large frequency detuning and high field intensities, the transition of atoms into the bound molecular state takes place almost non-oscillatory in time (only weakly expressed oscillations between the two population modes are observed). On the contrary, in the small detuning regime the hybrid atomic-molecular system displays large-amplitude Rabi-type oscillations between the populations. We illustrate the peculiarities of these two regimes using, along with the Landau-Zener model, several models with distinct properties, known from the linear theory of nonadiabatic transitions: the first Nikitin exponential-crossing model [6] that differs from the Landau-Zener case mainly by the finite final detuning at $t \to +\infty$, the first Demkov-Kunike quasi-linear level-crossing model of a finite pulse and finite detuning [7], and the Rosen-Zener finite-pulse constant-detuning, hence, non-crossing model [8]. Multiple level-crossing models are not considered.

Finally, we show that for some detuning modulation functions involving a finite final detuning (as it is the case for the first Nikitin model [6]) the strong coupling limit of high field intensities, perhaps surprisingly, is not optimal for molecule production.

The discussion below is restricted to the basic model of two-mode coherent molecule formation in an atomic Bose-Einstein condensate described by the following system of semiclassical mean-field Gross-Pitaevskii type coupled first-order time-dependent nonlinear differential equations [12-13]

$$i\frac{da_1}{dt} = U(t)e^{-i\delta(t)}\overline{a}_1 a_2, \quad i\frac{da_2}{dt} = \frac{U(t)}{2}e^{+i\delta(t)}a_1 a_1 \qquad (1)$$

that treat the atomic and molecular condensates as classical fields. Here $a_1$ and $a_2$ are the free-atomic and bound-molecular state amplitudes, respectively, $\overline{a}_1$ is the complex conjugate to $a_1$, the (real) functions $U(t)$ and $\delta(t)$ are characteristics of associating field – see below.

We suppose that all the quantities involved in the equations are dimensionless and that the initial conditions imposed are $|a_2(-\infty)|^2 = 0$, $|a_1(-\infty)|^2 = 1$, thus, we suppose that the system starts from an all-atomic state. The system possesses a motion integral indicating the conservation of the number of particles (number of atoms plus twice the number of molecules): $|a_1|^2 + 2|a_2|^2 = \text{const} = I_N$; we put here $I_N = 1$.

We do not discuss the approximations leading to this model as well as the ensuing applicability conditions since these questions have been extensively discussed from a number of points of view due to wide applicability of system (1) in diverse physical situations. For a detailed discussion of the limiting factors and the corresponding range of validity of equations (1) in cold molecule production via magnetically tunable Feshbach resonances see the recent review [14]. In the case of photoassociation more factors affect the mean-field dynamics described by system (1). The major limitations are caused by the spontaneous emission and rogue dissociation, i.e., formation of correlated atom pairs that do not belong to the molecular or atomic condensate modes. These aspects are analyzed, e.g., in [12,15-17].

Since the Feshbach resonance and the photoassociation are mathematically identical [12-14] and insight gathered in either case is applicable to the other, without loss of generality, we use here the photoassociation terminology. We thus refer to $U(t)$ as the Rabi frequency of the photoassociating laser field, and to $\delta(t)$ as the frequency modulation function defined as the time-integral of the laser frequency detuning from that of the atom-molecule transition. In the calculations below we operate exclusively with the derivative of $\delta(t)$ that we denote $\delta_t$ (hereafter we adopt the convention that the alphabetical index denotes differentiation with respect to the corresponding variable). A particular field configuration $\{U, \delta_t\}$ is then referred to as a model. The simplest possible model is the Rabi one for which the Rabi frequency and detuning are constant: $U = U_0 = \text{const}$, $\delta_t = \delta_0 = \text{const}$. This model is exactly solved both in linear and nonlinear cases [18,19]. Here we discuss the Landau-Zener: $U = U_0$, $\delta_t = 2\delta_0 t$ (Fig. 1a), the first Nikitin exponential-crossing: $U = U_0$, $\delta_t = \delta_0 (1 - e^{-at})$ (Fig. 1b), the first Demkov-Kunike: $U = U_0 \text{sech}(t)$, $\delta_t = 2\delta_0 \text{th}(t)$ (Fig. 2a), and the Rosen-Zener: $U = U_0 \text{sech}(t)$, $\delta_t = 2\delta_0$ (Fig. 2b) models. Though the Landau-Zener model is the prototype of all the term-crossing models, it describes a rather artificial process because of infinite-duration pulse and the diverging at infinity detuning the model deals with. More elaborate is the exponential-crossing model by Nikitin [6]. Here, the detuning is again an approximately linear function of time in the vicinity of the crossing point; however, at the end of the interaction it reaches a finite value. Since the pulse is again of infinite duration, it is expected that this model will incorporate the characteristics of both the Landau-Zener and Rabi problems. A model that has all the virtues of the Landau-Zener model and does not suffer its shortcomings is the first Demkov-Kunike model [7]. As it is seen from Fig. 2a, here the field pulse is of finite duration and the detuning is finite, hence, a Demkov-Kunike pulse models a physically realizable level-crossing process. As regards the Rosen-Zener model [8], it is a finite-pulse generalization of the Rabi non-crossing model. As such, this model serves as a standard reference for all non-crossing models.

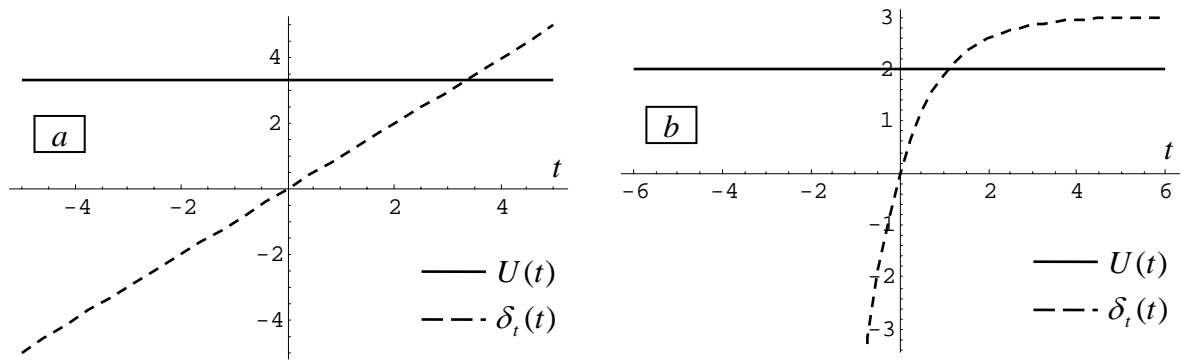

Fig. 1. a) Landau-Zener: $U = U_0$, $\delta_t = 2\delta_0 t$ and b) Nikitin exponential: $U = U_0$, $\delta_t = \delta_0 (1 - e^{-at})$ models.

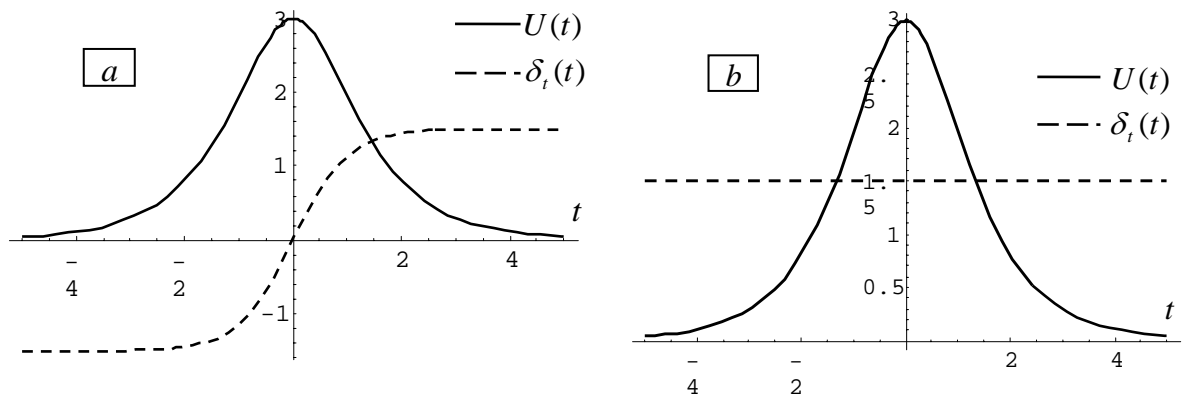

Fig. 2. a) Demkov-Kunike model: $U = U_0 \text{sech}(t)$, $\delta_t = 2\delta_0 \text{th}(t)$ and b) Rosen-Zener model: $U = U_0 \text{sech}(t)$, $\delta_t = 2\delta_0$.

As stated above, equations (1) describe the molecule formation processes both via photoassociation and through a Feshbach resonance. However, it should be noted that the two cases suggest different possibilities if considered from the experimental point of view because the coupling term $U$ cannot be adjusted in the case of a magnetic resonance [2,14]. So, the scenarios described by the models involving a time-dependent $U(t)$ do not apply for magnetic Feshbach resonances. This observation implies that the first Demkov-Kunike (Fig. 2a) and the Rosen-Zener (Fig. 2b) models apply to the photoassociation only. Hence, for the case of a Feshbach resonance one should discuss constant-$U$ modification of these models. In the Rosen-Zener case such a modification turns the model into the simple Rabi model [18] that has been discussed in detail in [19]. For the Demkov-Kunike model the constant coupling analogue, $U = U_0$, $\delta_t = 2\delta_0 \text{th}(t)$, is known to be the second Demkov-Kunike model. Evidently, this second model is the appropriate generalization of the Landau-Zener model for the Feshbach resonance case. This model seems to be the most feasible one from the point of view of experimental realization. For these reasons, below we discuss this model too.

We have previously shown [20] that the ***weak interaction limit*** in cold molecule formation is well treated using an exact nonlinear Volterra integral equation [21] for the molecular state probability. This equation allows the derivation of a highly accurate approximate solution for the case of small field intensities, $U_0^2 \leq 1$, using Picard's successive approximations [21]. The solution is written in terms of the solution to an auxiliary ***linear*** two-state problem thus indicating that the nonlinear evolution at weak coupling differs from the linear picture only quantitatively [20]. However, in the opposite limit of ***strong*** coupling, $U_0^2 \gg 1$, Picard's approximations, though remaining convergent, become almost useless in practice. For this reason, we appeal to the following exact nonlinear ordinary differential equation of the third order obeyed by the molecular state probability $p(t) = |a_2|^2$:

$$p_{ttt} - \left(\frac{\delta_{tt}}{\delta_t} + 2\frac{U_t}{U}\right)p_{tt} + \left[\delta_t^2 + 4U^2(1-3p) - \left(\frac{U_t}{U}\right)_t + \frac{U_t}{U}\left(\frac{\delta_{tt}}{\delta_t} + \frac{U_t}{U}\right)\right]p_t + \frac{U^2}{2}\left(\frac{\delta_{tt}}{\delta_t} - \frac{U_t}{U}\right)(1-8p+12p^2) = 0. \tag{2}$$

This equation is rather complicated; however, it is considerably simplified, as it is readily seen, in the case of constant field amplitude. Fortunately, system (1) for any model with varying field amplitude can be reduced to an equivalent system with constant Rabi frequency via the transformation of the independent variable

$$z(t) = \int_{t_0}^{t} \frac{U(t')}{U_0} dt' \qquad (3)$$

(a convenient choice here is $U_0 = \max[U(t)]$). This transformation changes equation (2) to the following much simpler form:

$$p_{zzz} - \frac{\delta_{zz}^*}{\delta_z^*} p_{zz} + \left[\delta_z^{*2} + 4U_0^2(1-3p)\right] p_z + \frac{U_0^2}{2} \frac{\delta_{zz}^*}{\delta_z^*} \left(1 - 8p + 12p^2\right) = 0, \qquad (4)$$

where the effective detuning $\delta_z^*$ is defined as

$$\delta_z^*(z(t)) = \delta_t(t) \frac{U_0}{U(t)}. \qquad (5)$$

The nonlinear terms in equation (4) are proportional to the field intensity $U_0^2$. Hence, the strong nonlinearity regime corresponds to high field strengths and we thus suppose that $U_0^2$ is a large parameter. Note now that the function $\delta_z^{*2}$ may also adopt large values (e.g., in the Landau-Zener model [5] $\delta_z^{*2} \sim z^2 = t^2$ diverges at $t \to \pm\infty$). Furthermore, note that the nonlinearity is merely determined by the current value of $p(t)$. Hence, at strong coupling the probability $p(t)$ should reach large values during the evolution of the system (of course, relatively large, since the normalization constraint $p$ cannot exceed $1/2$). Having these observations in mind, we suppose that the leading terms in equation (4) are the last two so that we neglect, for a while, the first two terms thus arriving at the following *limit* nonlinear equation of the first order:

$$\left[\delta_z^{*2} + 4U_0^2(1 - 3p_0)\right] p_{0z} + \frac{U_0^2}{2} \frac{\delta_{zz}^*}{\delta_z^*} \left(1 - 8p_0 + 12p_0^2\right) = 0. \qquad (6)$$

This is a productive equation. In spite of the singular way it was derived (the higher-order derivatives have been disregarded) the equation well works due to its rich structure that incorporates all the principal features of the exact equation (4), i.e., the form of the nonlinearity, the interplay between the nonlinearity and the detuning modulation, etc.

This equation has always trivial constant solutions $p_0 = 1/2$ and $p_0 = 1/6$ that are also stationary solutions to the exact equation (2). These solutions play, as we will see below, a pronounced role in determination of the asymptotes of the solution to system (1) at the strong coupling limit. Furthermore, the *general* solution to this equation, in spite of the complexity of the latter, is written in analytic form for arbitrary detuning modulation $\delta_z^*$ by a transformation involving interchange of dependent and independent variables. Indeed, it is possible to find such transformation of the independent variable $z \to s$ that reduces nonlinear equation (6) to a linear equation for the new variable $s$ if $s$ is considered as a **_dependent_** variable, $p_0$ serving then as the independent variable. This is achieved by choosing $s = U_0^2 / \delta_z^{*2}$. The resultant equation is written as

$$\left(p_0 - \frac{1}{6}\right)\left(p_0 - \frac{1}{2}\right)\frac{ds}{dp_0} + 4\left(p_0 - \frac{1}{3}\right)s - \frac{1}{3} = 0. \quad (7)$$

After simple integration we then arrive at the following principal result:

$$\frac{U_0^2}{\delta_z^{*2}} = \frac{C + p_0(p_0 - 1/2)^2}{9(p_0 - 1/6)^2(p_0 - 1/2)^2}, \quad C = \text{const}. \quad (8)$$

This algebraic equation, determining the limit form of the molecule formation probability at strong coupling $p_0(t)$ in terms of the time-variation of the effective frequency detuning $\delta_z^*(z(t))$, leads to several immediate conclusions. Indeed, note that if the detuning goes to zero at some point of the time the molecular state probability should inevitably adopt $1/6$ or $1/2$ at that point. Hence, the molecular probability is strictly equal [indeed, within the applicability limitations of the limit equation (6)] to $1/6$ at the frequency resonance crossing point. It then follows that for non-crossing models the molecular state probability cannot exceed $1/6$, hence, the sweep through the resonance is a necessary condition to create a considerable molecular population (recall that we start from pure atomic state).

A little examination shows that to define the constant $C$ in equation (8) the behavior of the function $\delta_z^{*2}$ at $t \to -\infty$ should be considered. It is not difficult to verify that for the first four particular models considered here, the Landau-Zener, the first Nikitin, the first Demkov-Kunike and the Rosen-Zener models [5-8], holds $\lim_{t \to -\infty}|\delta_z^*| = \infty$. Imposing now the initial condition $p_0(t = -\infty) = 0$ we get for these models $C = 0$. The situation is changed in the case of the second Demkov-Kunike model. Here, in principle, we have a non-zero constant, namely, $C = U_0^2/(64\delta_0^2)$. However, as shown below, the application of the limit equation (6) is in fact limited by the case of large detuning. Hence, we suppose $U_0^2/(64\delta_1^2) \ll 1$ and put $C \approx 0$ in this case as well. Furthermore, the zero integration constant means that the quartic equation (8) is reduced to a quadratic one. As a result, for all the discussed models we finally get

$$p_0 = \frac{1}{6} + \frac{\delta_z^*}{18U_0}\left(\frac{\delta_z^*}{U_0} - S\sqrt{\left(\frac{\delta_z^*}{U_0}\right)^2 + 6}\right), \quad S = \text{sgn}\left(\lim_{t\to-\infty}\delta_t(t)\right). \qquad (9)$$

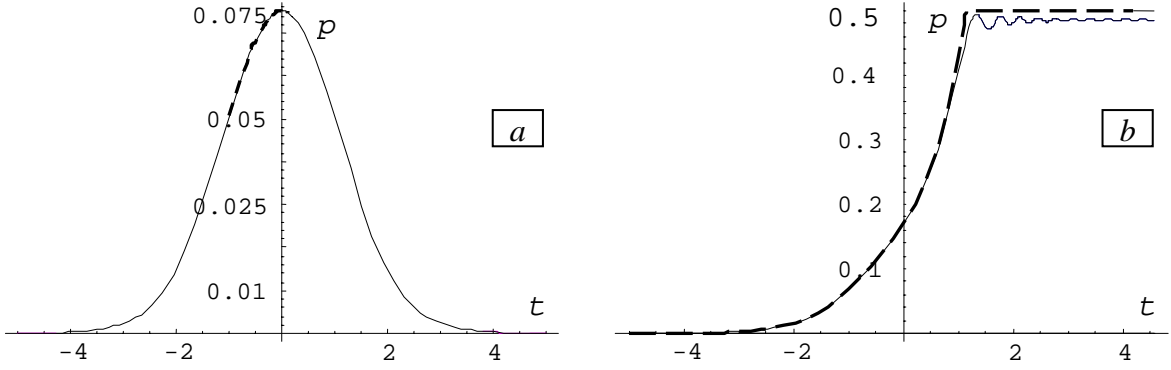

Fig. 3. Molecule formation probability vs. time ($U_0 = 20$, $\delta_0 = 10$). Solid line - numerical solution, dashed line - solution (9). a) Rosen-Zener model b) Demkov-Kunike model.

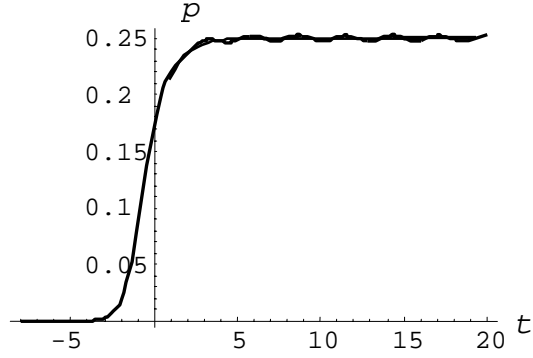

Fig.4. Nikitin exponential model, $U_0^2/\delta_0^2 = 4$. Monotonic curve represents solution (9).

The derived solution is a rather accurate approximation. This is demonstrated in Figures 3 and 4, where we compare this result with the numerical solution to Eq. (2). Note here that if the probability defined by this formula reaches, at a point $t = t_c$, the maximal value $1/2$ allowed by the normalization, it must be combined with the trivial solution $p_0 = 1/2$ for $t > t_c$. This happens, e.g., for the Landau-Zener [22] and the first Demkov-Kunike models (see Fig. 3b).

Furthermore, this solution allows the drawing of several qualitative conclusions of practical significance. First, suppose that the limit solution $p_0$ always remains less than $1/2$ or, equivalently, $\delta_t/U \leq \sqrt{2}$, if $S = -1$, and $\delta_t/U \geq -\sqrt{2}$, if $S = 1$. In this case if $\lim_{t \to -\infty} \delta_z^*(z(t)) = \lim_{t \to +\infty} \delta_z^*(z(t))$, then after interaction the system will **_return_** to its initial, pure atomic state. This happens, for instance, when the external field configuration is defined by the Rosen-Zener model (Fig. 3a). Note that the maximum molecular population achieved at this non-crossing process is less than $1/6$. Second, let in addition $\delta_z^*$ remain restricted in the neighborhood of $t = +\infty$ for any finite values of the detuning and Rabi frequency parameters. Then the final transition probability tends to $1/6$, when $U_0$ tends to infinity. This behavior occurs for the Nikitin exponential model (Fig. 4). This result proves that the application of high field intensities is not always effective to achieve large final molecular population (for more details on this point the reader is referred to reference [23] where this point is discussed in detail). The above general observations form the main physical result of the present paper.

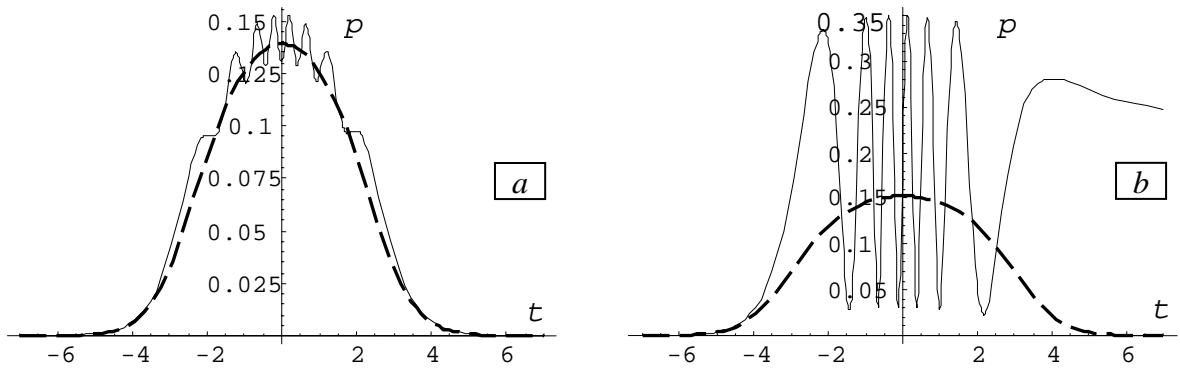

Fig. 5. Rosen-Zener model at small detuning regime, a) $U_0 = 10$, $\delta_1 = 1$, b) $U_0 = 10$, $\delta_1 = 0.2$. Solid line - numerical solution, dashed line - limit solution (9).

The common feature of the limit solutions for the five considered models is their non-oscillatory behavior. To find the conditions when such behavior is the case the approximate solutions for the Rosen-Zener, the two Demkov-Kunike and the Nikitin models are compared with the numerical results (the Landau-Zener case differs, see below). The analysis shows that the limit solution displays pronounced discrepancy as $\delta_0$ and $U_0$ are lowered (Fig. 5). Finally, we conclude that solution (9) is a good approximation when $U_0 >> 1$ and $\delta_0 > 1$.

Thus, to construct an approximate solution for the parameter variation range $U_0 >> 1$ and $\delta_1 < 1$ a different approach should be developed. The numerical examination shows that in this case the behavior of the system is much more "unstable": the time evolution of the molecular state probability reveals oscillations with large amplitude and varying frequency.

This behavior is qualitatively understood and a valid approximation for this regime is constructed by factorizing the exact third-order equation (4) as follows:

$$\left(\frac{d}{dz} - \frac{\delta_{zz}^*}{\delta_z^*}\right)\left[p_{zz} - \frac{U_0^2}{2}\left(1 - 8p + 12p^2\right)\right] + \delta_z^{*2} p_z = 0. \quad (10)$$

This form of the governing equation readily reveals the origin and the nature of different possible evolution scenarios of the association process. Indeed, as stated above, all the quantities involved in the equation are supposed to be dimensionless. As is seen from the formulas describing the time-dependence of the models, in our case this has been achieved by putting the time-scale equal to unity. It follows then that equation (10) involves two principal parameters: $U_0$ which characterizes the coupling and $\delta_0$ that stands for the magnitude of the detuning. Of course, the models may involve other parameters characterizing the field amplitude and the detuning modulation (e.g., in the Nikitin case we have an additional parameter $a$ that defines the resonance crossing speed). However, from the point of view of the structure of equation (10), these are secondary parameters, the principal two are $U_0$ and $\delta_0$. Furthermore, note that $\delta_0$ is not revealed in the term $\delta_{zz}^*/\delta_z^*$, hence, in the operator $(d/dz - \delta_{zz}^*/\delta_z^*)$. This suggests that these parameters do not interfere directly, their influence is well separated. It is then understood that different scenarios are due to the parameters' relative values compared with the time-scale, the unity. Indeed, consider, first, the case of small field intensities: $U_0 \ll 1$. Since the nonlinear term is proportional to $U_0$, it is understood that this regime, the week coupling case, should have much in common with the linear evolution. As shown in [20], this is, indeed, the case. By the same arguments, it is understood that the opposite limit of strong coupling, $U_0 \gg 1$, should always be strongly nonlinear. Now, if $\delta_0$ is also a large parameter the term $p_{zz}$ can be neglected and we arrive at above scenario described by the limit equation (6). The situation is changed if $\delta_0$ is small compared with the unity. It is expected that this time it is the turn to neglect the last term in equation (10). Basically speaking, this is the mathematical essence of the second strongly nonlinear scenario that we are in the position to analyze.

Before passing to this discussion, however, it should be noted that there is a remarkable exception when above speculations do not apply. This is the case of the Landau-Zener model. The point here is that in this case the detuning is proportional to the time, $\delta_t = 2\delta_0 t$, so that the time-scale and $\delta_0$ are not independent. Actually, in this case the time-scale is determined by $\delta_0$. To be precise, one has to consider the detuning modulation function $\delta = \int \delta_t dt = \delta_0 t^2$ and choose the time-scale as $T = 1/\sqrt{\delta_0}$. Applying this and dividing equation (10) by $\delta_0$ one convinces that in this particular case there exists only one independent parameter, the Landau-Zener parameter $\lambda = U_0^2 / \delta_0$. Thus, the second scenario does not apply to the Landau-Zener model – at strong coupling this model displays only one nonlinear evolution picture. Regarding the other case that involves unrestricted detuning, the first Nikitin exponential-crossing model, though it is rather similar to the Landau-Zener case, it does involve an independent parameter $\delta_0$ characterizing the magnitude of the detuning (more precisely, a final value of the detuning). For this reason, this model does reveal the second scenario of the strongly nonlinear evolution. The model is treated in detail in [23]. The results support the above observation that the secondary parameters, such as the parameter $a$, characterizing the resonance crossing speed, do not interfere much the two general scenarios of the nonlinear evolution.

Consider now the small detuning case in detail. Suppose that the detuning is restricted and the last term in the left-hand side of equation (10) is the smallest term of the equation at all time-points. Though it is expected that the role of this term should be rather restricted, it turns out that the term cannot be completely neglected even if one wants to construct a very rough, a zero-order, approximation. Indeed, disregarding the term leads to a monotonic solution

$$p = \frac{1}{2} \tanh^2 \left( \frac{U_0 z}{\sqrt{2}} \right) \qquad (11)$$

that does not possess the pronounced oscillatory feature displayed in Figure 5, i.e., the most important property of the evolution of the system in this regime. Thus, the limiting case $\delta_0 \equiv 0$ is degenerate. Physically, this degeneracy means that the exact resonance ($\delta_t = 0$) solution is unstable, hence, is not realizable experimentally because any perturbation will run the system far from the behavior described by solution (11). Actually, the last observation suggests a key to resolve the situation: one may try to introduce a (small) perturbation that is potent to lead to an appropriate zero-order solution. In fact, the last term merely plays this role, the role of a small perturbation. The idea is to replace the actual term by a perturbation that allows construction of an analytic solution. The easiest way to do this is to add and subtract a small constant (of the order of $\delta_0^2$ as $\delta_0$ goes to zero, i.e., presumably of the order of the term $\delta_z^{*2} p_z$) inside the square brackets in equation (10):

$$\left(\frac{d}{dz} - \frac{\delta_{zz}^*}{\delta_z^*}\right)\left[p_{zz} - \frac{U_0^2}{2}\left(1 - 8p + 12p^2\right) + A - A\right] + \delta_z^{*2} p_z = 0 \ . \qquad (12)$$

Now, the strategy is to neglect the last two terms, then to construct the general solution to the truncated equation with parameter $A$ being variable, and further to adjust the value of this parameter in such a way that the influence of the neglected terms calculated using the derived solution becomes minimal.

The described program is performed in a few straightforward steps since the truncated equation is identified with the equation obeyed by the Jacobi elliptic-sine function [24]. As a result, we arrive at the following approximation:

$$p = \left(\sqrt{p_1}\ \mathrm{sn}[\sqrt{p_2}\, U_0 (z - z_0); m]\right)^2, \qquad (13)$$

where

$$p_{1,2} = \frac{1}{2} \mp \sqrt{\frac{A}{2 U_0^2}}, \quad m = \frac{p_1}{p_2}. \qquad (14)$$

This is an oscillatory solution the behavior of which displays the needed qualitative features of the exact solution. The period of the oscillations is given as

$$T(m) = \frac{\pi}{\sqrt{p_2}\, U_0} \cdot {}_2F_1(1/2, 1/2; 1; m) \ . \qquad (15)$$

Furthermore, the numerical simulations show that there can always be found such a value of the parameter $A$ for which the approximate solution almost perfectly coincides with the numerical one (the graphs are practically indistinguishable). In order to find the appropriate value of $A$ one should examine the behavior of the neglected terms, namely,

$$R = A \frac{\delta_{zz}^*}{\delta_z^*} + \delta_z^{*2} p_z. \qquad (16)$$

Let us demonstrate the corresponding calculations using the example of the Rosen-Zener model. We have

$$\delta_z^* = \frac{2\delta_0}{\sin(z)}, \quad \frac{\delta_{zz}^*}{\delta_z^*} = -\frac{\cos(z)}{\sin(z)}, \qquad (17)$$

where $\qquad z = \frac{\pi}{2} + 2\,\mathrm{arctg}(\tanh(t/2)). \qquad (18)$

It is now immediately seen that the worst region is the vicinity of the point $z = 0$ where $R$ diverges (this point corresponds to the time-point $t = -\infty$, i.e., to the beginning of the interaction). An immediate suggestion is then to require $R$ to vanish at this point, thus removing the infinite perturbation. Since

$$p_0\big|_{z\to 0} \sim p_1 p_2 U_0^2 z^2, \tag{19}$$

this leads to the result

$$A = \frac{2\delta_0^2 U_0^2}{1+4\delta_0^2}. \tag{20}$$

Now, the comparison of this solution with the numerical one shows that the coincidence is good only for a few first oscillation cycles. Fortunately, the result is readily improved by slightly more elaborate speculations concerning the overall influence of the neglected terms. The final expression reads:

$$A \approx \sqrt{2}\,\delta_0^2 U_0^2, \tag{21}$$

so that the parameters $p_1$ and $p_2$ of the solution are given by a simple formula:

$$p_{1,2} \approx \frac{1}{2} \mp \frac{\delta_0}{\sqrt{\sqrt{2}}}. \tag{22}$$

The solution (13) is analogous to the nonlinear Rabi-solution [19]. As it is immediately seen, it is of universal form for arbitrary pulse shape and detuning modulation functions; the change of the laser field configuration only affects the argument and the expression for $A$ leaving the function itself unchanged. Moreover, depending on the particular model, the value of $A$ is varied in a very narrow interval. Hence, the qualitative behavior of the system in this regime is not much sensitive to the concrete form of the laser excitation. Another interesting feature, as follows from equation (15), is a subtle dependence of the oscillation frequency of the atom-molecule mixture on the parameters of the modulation of the laser field. In particular, the dependence of the frequency on the field amplitude is nonlinear, and moreover, oscillatory.

Thus, we have analyzed the strong interaction limit ($U_0 \gg 1$) of the two-mode one-color photoassociation of an atomic Bose-Einstein condensate and developed a general strategy for solving the problem for arbitrary pulse shape and detuning modulation functions. We have shown that there are two distinct scenarios of the system evolution - the large and small detuning regimes. The main peculiarity of the photoassociation process in the large detuning regime is its almost non-oscillatory behavior, i.e., only weak oscillations between the atomic and molecular populations occur. This regime is well approximated by the non-oscillatory limit solution (9). On the contrary, in the small detuning regime the evolution of the system is essentially oscillatory; in this case the approximate solution is expressed in terms of the oscillatory Jacobi sn-function [Eq. (11)]. The origin of the oscillations is qualitatively understood by examining the effective interaction time. If the final detuning is large, then in the regions relatively far from the crossing point, where the field amplitude is small, the interaction is rather weak, and the system does not change its state considerably. However, in the case of small detuning the effective interaction time is large, hence, during this time period the system will considerably change its state despite the smallness of the Rabi frequency: large-amplitude Rabi oscillations start.

The analysis presented above is essentially based on equation (4) [or Eq. (2)] for the molecular state probability. We have seen that this is a productive equation that allows one to considerably advance in the analysis of the solution to the initial system (1). However, it should be noted that the derivation of this equation is necessarily based on the application of the motion integral $|a_1|^2 + 2|a_2|^2 = \text{const} = I_N$ and the equation incorporates the adopted normalization $I_N = 1$. In other words, equation (4) essentially rests on the preservation of the particles. If the influence of the losses is discussed, this turns to be a decisive point because in that case the initial system of governing equations (1) and equation (2) become not equivalent. To this end, we may state the following. In general, the losses are modelled (within the mean-field approach adopted above) by adding an imaginary term in the detuning $\delta_t$. Hence, if one has constructed the general (complex) solution to the initial system (1), the losses can be treated simply by considering a complex detuning. However, it is impossible to construct the general complex solution using Eq. (4). Hence, more elaborate methods are needed. We have tried several approaches and the conclusion is that a trivial generalization of the above approach does not seem possible. We hope to discuss this problem in future.

This work was supported by the ISTC Grant N. A-1241, the Academy of Finland Project N. 115682, and the grant RA N. 104-2008.